\documentclass[journal=langd5,manuscript=article,layout=traditional]{achemso}

\usepackage{chemformula} 
\usepackage[T1]{fontenc} 
\usepackage{graphicx}
\usepackage{bm}
\usepackage{amsmath}

\author{Yixin Zhang}
\affiliation{School of Engineering, University of Warwick, Coventry CV4 7AL, United Kingdom}
\author{Duncan A. Lockerby}
\email{D.Lockerby@Warwick.ac.uk}
\affiliation{School of Engineering, University of Warwick, Coventry CV4 7AL, United Kingdom}
\author{James E. Sprittles}
\email{J.E.Sprittles@Warwick.ac.uk}
\affiliation{Mathematics Institute, University of Warwick, Coventry CV4 7AL, United Kingdom}

\title
  {Relaxation of Thermal Capillary Waves for Nanoscale Liquid Films on Anisotropic-slip Substrates}

\abbreviations{IR,NMR,UV}

\begin{document}


\begin{tocentry}

\includegraphics[scale=1]{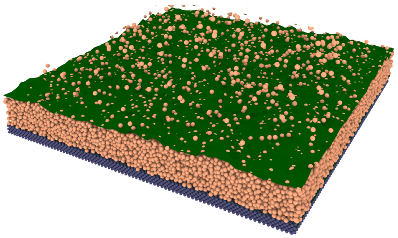}





\end{tocentry}

\begin{abstract}
The relaxation dynamics of thermal capillary waves for nanoscale liquid films on anisotropic-slip substrates are investigated, using both molecular dynamics (MD) simulations and a Langevin model. The anisotropy of slip on substrates is achieved using a specific lattice plane of a face-centred cubic lattice. This surface's anisotropy breaks the simple scalar proportionality between slip velocity and wall shear stress and requires the introduction of a slip-coefficient tensor. The Langevin equation can describe both the growth of capillary wave spectra and the relaxation of capillary wave correlations, with the former providing a time scale for the surface to reach thermal equilibrium. Temporal correlations of interfacial Fourier modes, measured at thermal equilibrium in MD, demonstrate that (i) larger slip lengths lead to a faster decay in wave correlations, and (ii) unlike on isotropic-slip substrates, the time correlations of waves on anisotropic-slip substrates are wave-direction dependent. These findings emerge naturally from the proposed Langevin equation, which becomes wave-direction dependent, agrees well with MD results, and allows to produce experimentally-verifiable predictions.
\end{abstract}

\section{Introduction}
The flourishing field of microfluidics and nanofluidics has encouraged many researchers to explore the nature of fluid flows at small scales, which is often qualitatively different from those at the macroscale\cite{bo2010}. For example, surface effects play an increasingly important role, as the surface-to-volume ratio increases. One of the most important surface effects is the presence of slip at the liquid-solid interface, which has been studied extensively by theoretical, numerical, and experimental methods\cite{la2005}. The breakdown of the no-slip condition, which is a foundation of fluid mechanics at the macroscale, manifests itself in physical phenomena in nanoflows and creates new engineering opportunities. One well-known effect is that slip can lead to very large flow-rate enhancement in micro- or nano-channels\cite{fa2010}, so that, for example, high-slip carbon nanotubes are promising membrane materials for ultra-filtration devices\cite{bo2018,da2014}.

Due to the importance of slip for fluid flows at small scales, seeking accurate methods to measure the slip length is necessary. However, this is challenging since such experiments often have to be performed at small scales where invasive techniques are complex. For example, known measurement methods (see the review\cite{la2005}) include: determining the flow rate at a given pressure drop\cite{ch2002}; measuring the drainage force felt by a submerged sphere approaching a solid using Surface Force Apparatuses and Atomic Force Microscopes\cite{chan1985,zhu2002,cr2001}; and finding the flow velocity near a solid directly using Particle Image Velocimetry (PIV)\cite{jo2005}, fluorescence recovery and fluorescence cross-correlations\cite{pi2000}. Notably, the aforementioned methods all have to impose a flow by external forces, which are difficult to control at small scales and can easily alter the properties of the underlying system, e.g. through introducing nanobubbles \cite{joly2006} or deforming soft/biological matter. Therefore, a non-invasive method to measure slip length is needed. Such a method, based on the measurement of the thermal motion of confined colloids, has been proposed and is able to achieve nanometric resolution of the slip length measurement and avoid generating nanobubbles\cite{joly2006}.

An alternative non-invasive technique involves investigating the thermally excited capillary waves of a liquid film at rest on a substrate\cite{ji2007,po2015,al2012,ki2003}. At thermal equilibrium, these waves can be described by the famous capillary wave theory\cite{aa2004,ma2017}. Recently, capillary wave theory has been extended using a Langevin equation to describe the dynamics of the nanowaves excited by thermal fluctuations and their approach to thermal equilibrium. Scaling relations are found for the process of surface roughening and the time scale for a smooth surface to reach thermal equilibrium is extracted\cite{zhang2020Thermal}. When in thermal equilibrium, it is well known that the temporal correlations of surface modes show an exponential decay, with a decay rate given by the dispersion relation of the system. Based on this, X-ray Photon Correlation Spectroscopy (XPCS) or similar techniques, can be used to infer the features of liquid-solid systems such as surface tension, viscoelasticity, and substrate surface structures\cite{ji2007,po2015,al2012,ki2003}. This technique has the advantage of being non-invasive since no external forces are applied.

 For thin liquid films, the dispersion relation for surface waves depends on the level of slip\cite{he2007} so that measuring the correlations of thermal capillary waves appears to be a potential method for inferring slip. In fact, this has been recently attempted in experiments for a hexadecane film on a glass\cite{po2015}. However, the negative slip length obtained is in conflict with the results from other more traditional (invasive) methods\cite{pi2000}. The conflicting result may be due to external forces used in the previous work\cite{pi2000} to drive the liquid, which lead to non-linear slip\cite{po2015}; in any case, it is clear that further research in this direction is required. Furthermore, the existing experimental study\cite{po2015} has not considered the exciting possibility that slip is anisotropic.

At larger than the nanoscale, it is commonplace to engineer anisotropic surfaces so that slip is flow-direction dependent, which can be used to control drag\cite{busse2012,woolford2009,fukagata2006theoretical}. 
Such surfaces can be created by patterning micro-ribs and micro-cavities onto a substrate and their effects on flow are measured using the PIV technique\cite{woolford2009}. However, at the nanoscale much less is known, partially due, as discussed, to the complexity and the potential problem of invasive measurement methods at these scales. Molecular dynamics simulations (MD) is a convenient tool to perform virtual experiments and investigate this problem. While there are previous MD studies of the relaxation of thermal capillary waves\cite{wi2010,th2008}, they did not consider the liquid-solid slip, even in the isotropic case.
 
In this work, we use MD simulations to investigate the relaxation of temporal correlations in thermal capillary waves for liquid nanofilms on isotropic and anisotropic-slip substrates. Different levels of slip are generated using different planes of a face-centred cubic lattice to form the substrates. The temporal correlations of surface modes are obtained from simulations after the free surface reaches thermal equilibrium, and they are compared with a Langevin model, which uses a newly derived dispersion relation that incorporates the anisotropic-slip length for the first time. The slip length used in the dispersion relation is directly measured from independent MD simulations. Thus, this work establishes the applicability and usefulness of using thermal capillary waves, as a non-invasive method, to measure anisotropic slip. 

This paper is organized as follows. We start with presenting molecular models to measure the relaxation of capillary wave correlations for liquid nanofilms on anisotropic-slip substrates, and then show the independent measurements of anisotropic slip length by a pressure-driven method. After that, in the theoretical section, we show the Langevin model of capillary wave dynamics and derive the dispersion relation for Stokes flow with anisotropic slip effects. Subsequently, we show the comparison of MD simulations with theoretical models and discuss the effects of anisotropic-slip boundary conditions. Finally, we conclude our findings and outline future directions of research.
\section{Simulation methodology} \label{sec2}
\subsection{MD models of liquid films on substrates}
\begin{figure}[h]
\includegraphics [width=\linewidth]{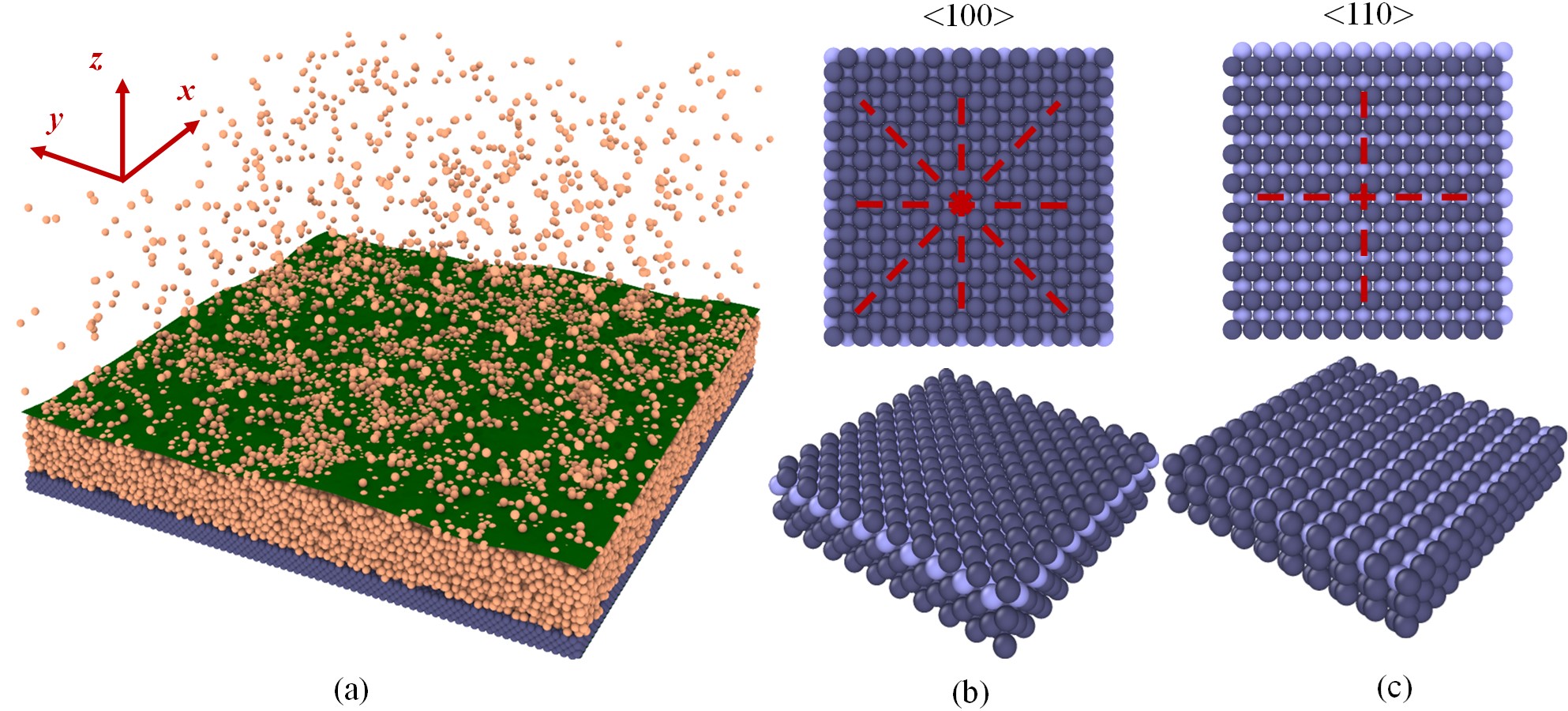}
\caption{\label{fig1} Snapshots of the set-up of a liquid film on a substrate in MD simulations. (a) A perspective view of the whole system. The fluid atoms are orange and the solid atoms are navy blue. The free surface generated by post-processing is coloured green. (b) Top view and perspective view of the isotropic $\left\langle 100 \right\rangle$ substrate surface. (c) Top and perspective view of the anisotropic $\left\langle 110 \right\rangle$ substrate surface. The red dash lines denote the lines of symmetry of the surface. The light blue color indicates solid atoms in the second layer.}
\end{figure}
Molecular dynamics simulations are used to simulate thermal capillary waves of liquid films on isotropic-slip and anisotropic-slip substrates. These simulations are performed in the open-source software LAMMPS\cite{pl1995}. The domain contains three phases with the liquid bounded by the vapor above and the solid below, as shown in Figure \ref{fig1}(a).
The liquid of the film is argon, simulated with the standard Lennard-Jones (LJ) 12-6 potential:
\begin{equation}
U\left({{r}_{ij}} \right)=4{{\varepsilon }_{ll}}\left[ {{\left(\frac{{{\sigma }_{ll}}}{{{r}_{ij}}} \right)}^{12}}-{{\left( \frac{{{\sigma }_{ll}}}{{{r}_{ij}}} \right)}^{6}} \right]
\end{equation}
where $r_{ij}$ is the pairwise distance. For argon, the energy parameter $\varepsilon_{ll}$ ($ll$ denotes liquid-liquid interactions), the length parameter $\sigma_{ll}$, and an atomic mass are $1.67\times{10^{-21} }$  J, 0.34 nm, and $6.63\times{10^{-26}}$ kg, respectively.  The temperature of this system is kept at $T=85$ K or $T^*=0.7 \varepsilon_{ll} /k_B$ (* henceforth denotes LJ units  and $k_B$ is the Boltzmann constant). At this temperature, the mass density of liquid argon is $1.40\times{10^{3}}$ kg/m\textsuperscript{3} and number density $n_{l}^*=0.83/\sigma_{ll}^3$. The number density of the vapor phase is about $(1/400) n^*$. As such, in the theoretical section of this article, the vapor is assumed to be dynamically passive and has no effect on the dynamics of capillary waves.

The substrate is platinum with a face-centred cubic (fcc) structure. It has been noticed before that the different planes of the fcc lattice lead to varying slip properties due to the difference in interfacial atom structures\cite{so2007}, see Figures \ref{fig1}(b) and (c). For example, the $\left\langle 100\right\rangle$ surface has four lines of reflectional symmetry, and so is for practical purposes isotropic: a unidirectional flow in the $x$ or $y$ direction generates the same slip velocity at the hydrodynamic boundary, see Figure \ref{fig2}(a). However, the $\left\langle 110\right\rangle$ surface is evidently anisotropic with only two lines of symmetry, since these unidirectional flows generate different slip velocities, see Figure \ref{fig2}(b). Therefore, in this work, we use 
the $\left\langle 100\right\rangle$ surface to represent a good approximation to an isotropic-slip substrate and the $\left\langle 110\right\rangle$ surface to represent an anisotropic-slip substrate. We note that the slip anisotropy of the $\left\langle 110\right\rangle$ surface was first discovered in the reference\cite{so2007}. However, their description of the resulting macroscopic slip boundary condition is incorrect, see the discussion in theoretical section.

The platinum mass density is $21.45\times{10^3}$  kg/m\textsuperscript{3} with an atomic mass of $3.24\times{10^{-25}}$ kg. The solid substrate is assumed to be rigid, which saves considerable computational cost compared to simulating flexible walls. In any case, previous studies show that the two kinds of walls produce very similar results\cite{fa2010}. The liquid-solid interactions are modelled by the same 12-6 LJ potential with $\varepsilon_{ls} =C\varepsilon_{ll}$ and $\sigma_{ls}=0.8\sigma_{ll}$. We vary $C$ to obtain different amounts of slip. For the $\left\langle 100\right\rangle$ surface, we choose case 1: $C_{\left\langle 100\right\rangle,\,1}=0.7$ and case 2: $C_{\left\langle 100\right\rangle,\,2}=0.35$. For the $\left\langle 110\right\rangle$ surface, we choose case $C_{\left\langle 110\right\rangle}=0.7$. The choice of the value of $C$ for all cases creates high wettability of the liquid on the solid. The cut-off distance, beyond which the intermolecular interactions are omitted, is chosen as ${r_c}^*=5.5 \sigma_{ll}$.

In order to compare with the predictions of a continuum model, the transport properties of liquid argon are calculated by simulating the equilibrium motion of a thick layer of liquid argon. It is found that the surface tension $\gamma = 1.52\times{10^{-2}} $ N/m (obtained from the virial expression) and the dynamic viscosity $\mu=  2.87\times{10^{-4}}$ kg/(ms) (obtained from the Green-Kubo relation). 

The initial dimensions of the liquid film $(L_x, L_y, h_0)$ in Figure \ref{fig1}(a) are chosen as, $L_x=31.4$ nm, $ L_y = 31.4$ nm, and $h_0=3.14$ nm.  The lateral size of the substrate is the same as that of the liquid film and it has a thickness $h_s=0.78$ nm.

We initialise each MD simulation as follows. The liquid film and vapour are equilibrated separately in periodic boxes at $T^*=0.7\varepsilon_{ll} /k_B$. The liquid film is then deposited above the substrate and the vapour on top of the film. Because there exists a gap (a depletion of liquid particles) between the solid and liquid, arising from the repulsive force in the LJ potential, it is necessary to deposit the liquid above the substrate by some distance. The thickness of the gap is found to be approximately $0.2$ nm at equilibrium, so that we choose a deposit distance $d=0.2$ nm. This makes the initial position of the film surface at $h_0+d=3.34$ nm. After assembly, the positions and velocities of the liquid and vapour atoms are updated with a Nos\'{e}-Hoover thermostat. Periodic boundary conditions are applied in the $x$ and $y$ directions of the system whilst vapour particles are reflected specularly in the $z$ direction at the top boundary of the system.

The free surface position is defined as the usual equimolar surface. The way to extract the surface profile $h(x,y,t)$ from MD simulations is detailed in our previous work\cite{zhang2019}. Since the density of vapour is much smaller than the liquid’s, and we are well away from the critical point, the position of the interface can be unambiguously defined. Then two-dimensional Fourier transforms of the surface profile are performed to obtain the amplitude of interfacial Fourier modes.

\subsection{Measurements of slip length} \label{sec3}
\begin{figure}
\includegraphics[width=\linewidth]{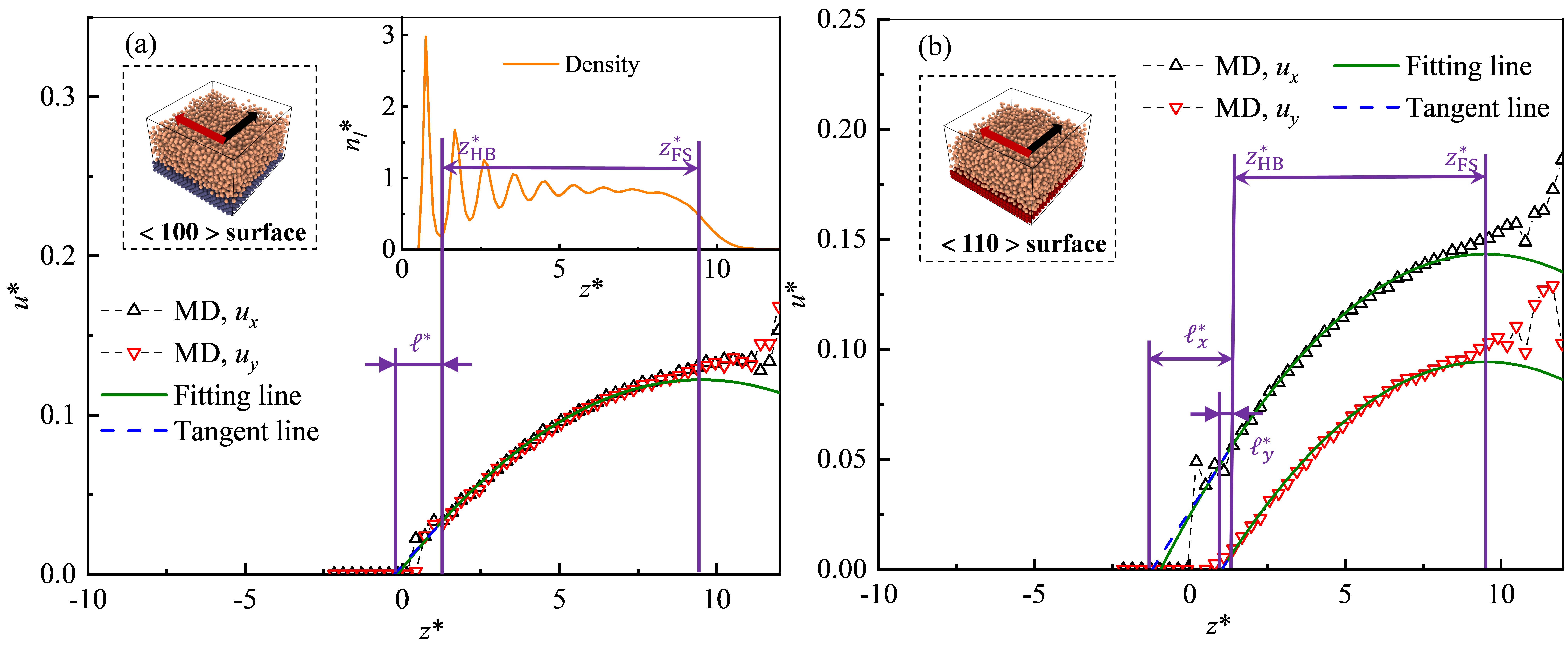}
\caption{Slip length measured using a pressure-driven flow past a plate. MD velocity (triangles) are fitted with analytical solutions (green solid lines) with the HB ($z_\mathrm{HB}$) at the first valley of MD density (orange solid line) and FS ($z_\mathrm{FS}$) at $0.5n_l^*$. The tangent line (blue dash lines) gives the slip length $\ell^*$. (a) for the isotropic $\left\langle 100\right\rangle$ surface; (b) for the anisotropic $\left\langle 110\right\rangle$ surface.}  \label{fig2}
\end{figure}

Slip length is measured from independent configurations by simulating unidirectional ($x-$ or $y-$direction) pressure-driven flows past a solid surface (Poiseuille flows) as shown by the MD snapshots in the top-left corner of Figures \ref{fig2}(a) and (b). The pressure gradient is created by applying a body force $g$ (along the $x$ or $y$ direction) to each particle in the fluid.
The generated velocity distribution is then parabolic: $u_{x,y}(z)=\frac{\rho g_{x,y}}{2\mu }(z-{{z}_\mathrm{HB}})(2{{z}_\mathrm{FS}}-z_\mathrm{HB}-z)+{{u}_{s}}$. Here $z_\mathrm{HB}$ and $z_\mathrm{FS}$ are positions of the hydrodynamic boundary (HB) and free surface (FS), respectively, and $u_s$ is the slip velocity at the HB. 

The precise location of two boundary positions is not trivial since there is an interfacial zone between the two different phases (solid-liquid and liquid-vapour) as shown in the density distribution (the orange line in Figure \ref{fig2}(a)). For the HB, research has shown it is located inside the liquid, between first-peak density and second-peak density rather than simply at the solid surface\cite{bo1994,ch2015} (an offset which matters when the interfacial layer has comparable width to the film). In line with this finding, we choose the position of HB at the first valley of density distribution: $z_\mathrm{HB}^*=1.2\sigma$ (see Figure \ref{fig2}(a)). The position of the FS is determined in the standard way by the location of equimolar surface where density is $0.5n_l^*$, with $z_\mathrm{FS}^*=9.5\sigma$ (see Figure \ref{fig2}(a)). After locating the boundary, the slip velocity is obtained by fitting velocity profiles of MD data (symbols) with analytical expressions of velocity (solid green lines) as shown in Figure \ref{fig2}. The slip length $\ell$ is the distance between the HB and the position where the the linear extrapolation of the velocity profile vanishes, see the dashed blue lines (since the slip length is small for each case, the tangent line is close to the fitting line).

Figure 2(a) is for the measurement of slip length of case $C_{\left\langle 100\right\rangle,\,1}$. The individually applied body forces are $g_x^*=0.01$ and $g_y^*=0.01$. It can be seen that the velocity profiles $u_x$ and $u_y$ are nearly the same, indicating that the $\left\langle 100\right\rangle$ surface is isotropic in terms of slip. The slip length is thus characterised by a scalar value $\ell^*=1.41\sigma$ (0.48 nm). In the same way, the slip length $\ell=3.34$ nm for case $C_{\left\langle 100\right\rangle,\,2}$ is obtained. We note that applying larger body forces may lead to a shear-dependent slip length\cite{zhang2020}, but here we have ensured that we remain in the regime where the measured slip length is constant independent of driving force magnitude.

For the $\left\langle 110\right\rangle$ surface, the individually applied body forces are also $g_x^*=0.01$ and $g_y^*=0.01$. There is a clear difference between $u_x$ and $u_y$ (see Figure \ref{fig2}(b)), indicating that the $\left\langle 110\right\rangle$ surface exhibits significant anisotropy. The slip lengths in the $x$ and $y$ direction are measured to be $\ell_x=0.80$ nm and $\ell_y=0.05$ nm, respectively, for the case $C_{\left\langle 110\right\rangle}$.

Notably, as the HB does not align with the edge of the solid, the effective thickness of the fluid domain simulated for capillary waves is different from its initial thickness. As the position of the initial free surface is at $3.34$ nm and the HB is at $z_\mathrm{HB} = 0.41$ nm, the effective thickness of a film is 2.93 nm. We note that our choice of the position of HB is empirical, and guided by the results of Bocquet and Barrat\cite{bo1994}; for a more robust determination of the HB, one may refer to refs\cite{bo1994,ch2015}.

\section{Theoretical approach}
\subsection{Langevin model for capillary wave dynamics}\label{sec4}
At thermal equilibrium, the static spectrum $S_s$ of thermal capillary waves is given by Capillary Wave Theory\,\cite{zhang2020,delgado2008hydrodynamics,th2008} , and for a two-dimensional surface, it is
\begin{equation}\label{ss}
{{S}_{s}}=\sqrt{{{\left\langle {{\left| \widehat{\delta h}\left(q\right) \right|}^{2}} \right\rangle }_{s}}}=\sqrt{{{L}^{2}}\frac{{{k}_{B}}T}{\gamma {{q}^{2}}}},
\end{equation}
where $\widehat{\delta h}(q)$ is the Fourier transform of height deviations (from their average height), wavenumber $q$ is the magnitude of a wave vector $\bm{q}=\left(q_x,q_y\right)$ with $q=\sqrt{q_x^2+q_y^2}$, $k_B$ is the Boltzmann constant, $T$ is temperature, and $\gamma$ is the surface tension. The $\langle\, \cdots\rangle$ denotes an ensemble average over time when the free surface is in thermal equilibrium, and $|\cdots|$ the norm of the transformed variable. Notably, the form of the Fourier transform adopted here is $\widehat{\delta h}(q)=\int\int\delta h(x,y)e^{-(iq_xx+iq_yy)}\,dx\,dy$, which makes $L^2$ in eq \ref{ss} appears as a numerator, in contrast to some previous works where it appears in the denominator due to a different definition for the transform\cite{delgado2008hydrodynamics,th2008}. It is also worth mentioning that the interface thickness will diverge logarithmically with the $L/D$ (D is a molecular cut-off length) based on the above static spectrum. The introduction of gravity or a certain binding potential, however, can remedy this problem \cite{rowlinson2013molecular} but it is not our concern in this article.
  
It is well-known that the temporal correlations of capillary waves measured in thermal equilibrium usually decay exponentially to zero (overdamped)\cite{ki2003,wi2010,th2008}
\begin{equation}\label{eqe}
\left\langle  \widehat{\delta h}\left(q,t\right){{\widehat{\delta h}}^{*}}\left(q,t'\right) \right\rangle =S_s^2\,{{e}^{-\Omega \left(q\right)\left| t-t' \right|}},
\end{equation}
where the asterisk denotes conjugate values. The decay rate $\Omega$ is given by the dispersion relation of the system (temporal growth rate of a surface mode). For a thin film with isotropic slip, the dispersion relation is\citep{oron1997long,zhang2019}
\begin{equation}\label{eqLEO}
\Omega(q)=\frac{\gamma}{3\mu}\left(h_0^3+\ell h_0^2\right)q^4,
\end{equation} 
which can be obtained from a linear stability analysis of a lubrication equation (assuming $2\pi h_0/L\ll 1$)\citep{oron1997long,zhang2019}
\begin{equation} \label{eqLE}
\frac{\partial h}{\partial t}=-\frac{\gamma}{\mu }\nabla \cdot \left[\left(\frac{1}{3}h^3+\ell h^2\right)\nabla  \nabla^2 h \right].
\end{equation}
Here $\nabla$ is the usual gradient operator. 

Obviously, one can derive a similar lubrication equation considering anisotropic slip and obtain the corresponding dispersion relation for the prediction of relaxation of capillary waves on anisotropic-slip substrates. However, the system simulated in this article has $2\pi h_0/L\approx 0.6$ so that the lubrication approach is not adopted for our work. Instead, we perform a linear stability analysis of Stokes flow and obtain the anisotropic dispersion relation $\Omega(q_x,q_y)$, without the need to assume $2\pi h_0/L\ll 1$. This process is presented in the next subsection.

In this subsection, we need to address the problem due to the assumption of thermal equilibrium in eq \ref{eqe}, since we do not know how long it will take for a free surface to reach its thermal equilibrium.
 
The dynamics of thermal capillary waves can be described by a Langevin equation, where the motion of an interfacial Fourier mode $\widehat{\delta h}$ is governed by the surface tension pressure $\gamma q^2 \widehat{\delta h}$ and a fluctuating pressure $\zeta \widehat{N}$\cite{zhang2020Thermal,me2005,davidovitch2005spreading,pi2016,zwanzig2001nonequilibrium}:
\begin{equation}
\frac{\gamma q^2}{\Omega(q_x,q_y)}\frac{\partial }{\partial{t}}\widehat{\delta h}=-\gamma q^2 \widehat{\delta h}+\zeta \widehat{N},
\end{equation}
where $\widehat{N}(\bm{q},t)$ is a complex Gaussian random variable with zero mean and covariance $\langle | \widehat{N}\widehat{{N'}}| \rangle=\delta( \bm{q}-\bm{q}')\delta(t-t')$, and $\zeta$ is the noise amplitude. The noise amplitude $\zeta$ can be found from the static spectra, as described in the previous work\cite{zhang2020Thermal}, so that: 
\begin{equation}\label{df}
\zeta= \, \sqrt{\frac{2}{\Omega}}\gamma q^2 \, S_\mathrm{s}\,.
\end{equation}
From the Langevin equation, the height-height correlation is found to be (using the It\^{o} integral\cite{di2016,me2005}) 
\begin{align}\label{eq4}
  \left\langle  \widehat{\delta h}({{q}_{x}},{{q}_{y}},t){{\widehat{\delta h}}^{*}}({{q}_{x}},{{q}_{y}},t')  \right\rangle =&\left\langle {{\left| \widehat{\delta h}({{q}_{x}},{{q}_{y}},0) \right|}^{2}} \right\rangle {{e}^{-\Omega ({{q}_{x}},{{q}_{y}})(t+t')}} \nonumber \\ 
 & -{{L}^{2}}\frac{{{k}_{B}}T}{\gamma {{q}^{2}}}\left[ {{e}^{-\Omega ({{q}_{x}},{{q}_{y}})(t+t')}}-{{e}^{-\Omega ({{q}_{x}},{{q}_{y}})\left| t-t' \right|}} \right]\,.
\end{align}
Notably, a similar expression can be obtained by a linear stability analysis of stochastic lubrication equations\cite{di2016,me2005,zhao2019,zhang2020,fe2007,zhang2020Thermal} instead of using the Langevin model presented here. However, as we have discussed, the lubrication approach is not suitable for our system.

Equation \ref{eq4} can describe two important aspects of capillary wave dynamics. The first aspect is the growth of capillary wave spectra to the static spectra (using $t=t'$ to obtain the equal-time correlations), namely, the process of surface roughening. For this case, we can assume that the free surface is initially smooth, $\langle {{| \widehat{\delta h}({{q}_{x}},{{q}_{y}},0) |}^{2}} \rangle=0$, so that eq \ref{eq4} is simplified to
\begin{equation}\label{eqs}
\left\langle {{\left| \widehat{\delta h}({{q}_{x}},{{q}_{y}},t) \right|}^{2}} \right\rangle ={{L}^{2}}\frac{{{k}_{B}}T}{\gamma {{q}^{2}}}\left[ 1-{{e}^{-2\Omega ({{q}_{x}},{{q}_{y}})t}} \right].
\end{equation}
Thus, the characteristic time scale for a smooth surface to reach the static spectra, i.e. thermal equilibrium, is $t_s=\mathrm{max}\left[1/\Omega(q_x,q_y)\right]
$. In the following, we use $S\left(q_x,q_y,t\right)=\sqrt{\langle {{| \widehat{\delta h}({{q}_{x}},{{q}_{y}},t) |}^{2}}\rangle}$ for notational simplicity.

The second aspect is the aforementioned relaxation of capillary wave correlations after a free surface reaches its static spectra (i.e. thermal equilibrium). In this case, the initial condition is $\langle {{| \widehat{\delta h}({{q}_{x}},{{q}_{y}},0) |}^{2}} \rangle=S_\mathrm{s}^2=L^2k_BT/\left(\gamma q^2\right)$ so that eq \ref{eq4} becomes
\begin{equation}\label{eq6}
\left\langle  \widehat{\delta h}({{q}_{x}},{{q}_{y}},t){{\widehat{\delta h}}^{*}}({{q}_{x}},{{q}_{y}},t')  \right\rangle ={{L}^{2}}\frac{{{k}_{B}}T}{\gamma {{q}^{2}}}{{e}^{-\Omega ({{q}_{x}},{{q}_{y}})\left| t-t' \right|}}.
\end{equation}
Thus, the Langevin equation is a unifying model, able to describe both the growth of capillary wave spectra towards thermal equilibrium and, once there, the relaxation of capillary wave correlations, with the former providing a time scale for the surface to reach thermal equilibrium. We note that eq \ref{eq6} can be derived alternatively using fluctuation-dissipation theorem\cite{he2007}, but this approach is unable to predict the growth of capillary waves at the same time. Also, the work in ref \cite{he2007} only considers isotropic-slip effects so that the dispersion relation in eq \ref{eq6} is $\Omega(q)$ instead of $\Omega(q_x,q_y)$. 

In the following, we define a dimensionless variable:
\begin{equation}\label{eqR}
R_{h_qh_q^*}\left(q_x,q_y,\left|t-t'\right| \right)=\left\langle  \widehat{\delta h}({{q}_{x}},{{q}_{y}},t){{\widehat{\delta h}}^{*}}({{q}_{x}},{{q}_{y}},t')  \right\rangle/\left({{L}^{2}}\frac{{{k}_{B}}T}{\gamma {{q}^{2}}}\right) 
\end{equation}
to normalise eq \ref{eq6} with the static spectrum. Note that in MD simulations, the temporal correction eq \ref{eqR} is calculated based on the average of the trajectories of a single simulation at thermal equilibrium. As we have shown theoretically from the Langevin equation, it is only in this case that the temporal correlations obtained will have the form of a simple exponential decay.
\subsection{Anisotropic boundary condition and dispersion relation}    
To derive the required dispersion relation, we perform a linear stability analysis of three-dimensional Stokes flow. We outline the governing equations for this problem and keep the details of the derivations in the Supporting Information. The liquid is assumed to be incompressible and the vapour is dynamically passive. Incompressibility requires that
\begin{equation}
\nabla \cdot \mathbf{u}=0,
\end{equation}
where $\mathbf{u}=(u_x,u_y,u_z)$, and $u_x,u_y,u_z$ are the velocities in $x,y,z$ directions, respectively. The momentum equation with the assumption of Stokes flow, is as follows
\begin{equation}
\mu\nabla^2\mathbf{u}=\nabla p,
\end{equation} 
where $\mu$ is the liquid viscosity and $p$ is the liquid's pressure with respect to that of the vapour. For the boundary conditions, at the position of free surface $z=h(x,y,t)$, we have the dynamic condition:
\begin{equation}
\bm{\tau}\cdot\mathbf{n}=-\left(\gamma\nabla\cdot\mathbf{n}\right)\mathbf{n},
\end{equation}
where $\bm{\tau}$ is the hydrodynamic stress tensor, $
\tau_{ij}=-p\delta_{ij}+\mu\left(\partial u_i/\partial x_j+\partial u_j/\partial x_i\right)$;
 $\gamma$ is the surface tension; and $\mathbf{n}$ is the outward normal to the free surface:
\begin{equation}
\mathbf{n}=\frac{\left( -\partial h/\partial x,-\partial h/\partial y,1 \right)}{\sqrt{1+{{\left( \partial h/\partial x \right)}^{2}}+{{\left( \partial h/\partial y \right)}^{2}}}}.
\end{equation}
Under the assumption of small perturbations ($\frac{\partial h}{\partial x},\frac{\partial h}{\partial y}\ll 1$), the dynamic boundary condition is reduced to (in the normal direction):
\begin{equation}
-p+\mu\frac{\partial u_z}{\partial z}=\gamma\left(\frac{\partial^2 h}{\partial x^2}+\frac{\partial^2 h}{\partial y^2}\right),
\end{equation}
and in the tangential directions to the surface
\begin{align}
&\frac{\partial u_x}{\partial z}+\frac{\partial u_z}{\partial x}=0\,,\\
&\frac{\partial u_y}{\partial z}+\frac{\partial u_z}{\partial y}=0\,.
\end{align}
The kinematic condition at the free surface is given by
\begin{equation}
u_z=\frac{\partial h}{\partial t}+u_x\frac{\partial h}{\partial x}+u_y\frac{\partial h}{\partial y}.
\end{equation}
At the substrate surface, the no penetration condition is,
\begin{equation}
u_z=0,
\end{equation}
and the anisotropic-slip boundary condition is, 
\begin{align}
& \label{eq16}u_x=\ell_x\frac{\partial u_x}{\partial z}\,,\\ 
&u_y=\ell_y\frac{\partial u_y}{\partial z}\,,\label{eq17}
\end{align} 
as we will explain now.

The Navier-slip boundary condition describes the proportionality of wall shear stress ($\tau$) to velocity slip ($u$) so that for a two-dimensional flow (and one-dimensional boundary) in the $x$-direction we have:
\begin{equation} \label{eq18}
u_x={\beta}\tau_x \, ,
\end{equation}
where $\beta=\ell/ \mu$. Extending eq \ref{eq18} to two-dimensional surfaces is straightforward when the surface is isotropic (i.e. when $\beta$ is scalar):
\begin{equation} \label{eq19}
\mathbf{u_{||}}=\beta \bm{\tau_{||}} \, .
\end{equation}
Here the subscript $||$ after a vector denotes its parallel components to the solid surface. At first thoughts, one could consider incorporating anisotropy into the Navier-slip condition by maintaining the form of eq \ref{eq19}, but making the slip length dependent on direction (e.g. $\beta(\theta)$)\cite{so2007}. However, this is incorrect, since it assumes that velocity slip is always in the direction of the wall shear stress. Consider, for example, the case $\beta_y=0$ (i.e. no-slip in this direction) and $\beta_x>0$: the $y$-component of slip ($u_y$) is zero irrespective of the direction of the wall shear stress, and therefore slip velocity and shear stress can easily be misaligned (apart from when $\bm{\tau_{||}}$ is in the $x$ direction). 

In other words, surface anisotropy breaks the simple scalar proportionality described by eq \ref{eq19}. To generalise the boundary condition requires the introduction of a slip-coefficient (second-order) tensor:
\begin{equation} \label{eq20}
\mathbf{u_{||}}=\bm{\beta}\cdot\bm{\tau_{||}} \, .
\end{equation}
For isotropic surfaces $\bm{\beta}=\beta\bm{I}$, where $\bm{I}$ is the identity tensor. In this article we consider slip on \emph{orthotropic} surfaces; surfaces that have two axes of symmetry (see Figure 1(c)), which are orthogonal (unlike isotropic surfaces, which have an infinite number). In the case of orthotropic surfaces:
\begin{equation}\label{eq21}
 \bm{\beta}=\begin{pmatrix}
\beta_1 & 0 \\
0 & \beta_2
\end{pmatrix}\, ,
\end{equation}
where $\beta_1$ and $\beta_2$ are slip related parameters along the two lines of symmetry (e.g. with the grain and against the grain). Thus, the slip boundary for the $\left\langle 110\right\rangle$ is obtained as eq \ref{eq16} and eq \ref{eq17}. 

Notably the slip boundary condition obtained by combining eq \ref{eq20} and eq \ref{eq21} assumes that the coordinate system is aligned to the axes of symmetry of the orthotropic surface. More generally, if this is not the case, we would have:
\begin{equation}
 \bm{\beta}=\begin{pmatrix}
\beta_x & \beta_{xy} \\
\beta_{yx} & \beta_{y}
\end{pmatrix}=\begin{pmatrix}
\beta_1 \cos{\alpha} & \beta_2 \sin{\alpha} \\
-\beta_1 \sin{\alpha} & \beta_2 \cos{\alpha}
\end{pmatrix}\, ,
\end{equation}
where $\alpha$ is the angle between the cartesian coordinate system and the orthotropic axes, but henceforth we do not consider this possibility as for our case we know $\alpha=0$.
 
It is also worth noting that the anisotropic slip considered in this work is fundamentally different from the heterogeneous slip where $\ell=\ell(x,y)$\cite{pi2016}, and the latter has already been incorporated in the relaxation dynamics of capillary waves by Pierre-Louis\cite{pi2016}.

With above eqs 7-17, the linear stability analysis gives the dispersion relation (see Supporting Information):
\begin{align}\label{dis}
  & \Omega \left( {{q}_{x}},{{q}_{y}} \right)=\frac{\gamma q}{4\mu }\frac{\sinh \left( 2q{{h}_{0}} \right)-2q{{h}_{0}}+4A{{\sinh }^{2}}\left( q{{h}_{0}} \right)}{{{q}^{2}}h_{0}^{2}+{{\cosh }^{2}}\left( q{{h}_{0}} \right)+A\left[ 2q{{h}_{0}}+\sinh \left( 2q{{h}_{0}} \right) \right]}, \nonumber\\ 
 & A=q{{\ell }_{x}}-q\frac{{{\ell }_{x}}-{{\ell }_{y}}}{1+\frac{q_{x}^{2}\left[ q{{\ell }_{y}}\sinh \left( q{{h}_{0}} \right)+\cosh \left( q{{h}_{0}} \right) \right]}{q_{y}^{2}\left[ q{{\ell }_{x}}\sinh \left( q{{h}_{0}} \right)+\cosh \left( q{{h}_{0}} \right) \right]}}. 
\end{align}
If $\ell_x =\ell_y$, eq \ref{dis} is reduced to the existing dispersion relation for films on isotropic substrates\cite{he2007,pi2016,ka2004}.

\section{Results and discussions} \label{sec5}
In this section, we present and discuss our MD simulation results and their comparison to analytical solutions. Firstly, we show the transient growth of capillary wave spectra to the static spectrum. Secondly, using the isotropic-slip $\left\langle 100\right\rangle$ substrate, we explore the effects of different slip length on the relaxation of capillary wave correlations. Thirdly, the effects of anisotropic slip on the relaxation of capillary wave correlations are demonstrated using the anisotropic $\left\langle 110\right\rangle$ substrate.
 
\subsection{Transient growth of capillary wave spectra}
\begin{figure}[t!]
\includegraphics[scale=0.1]{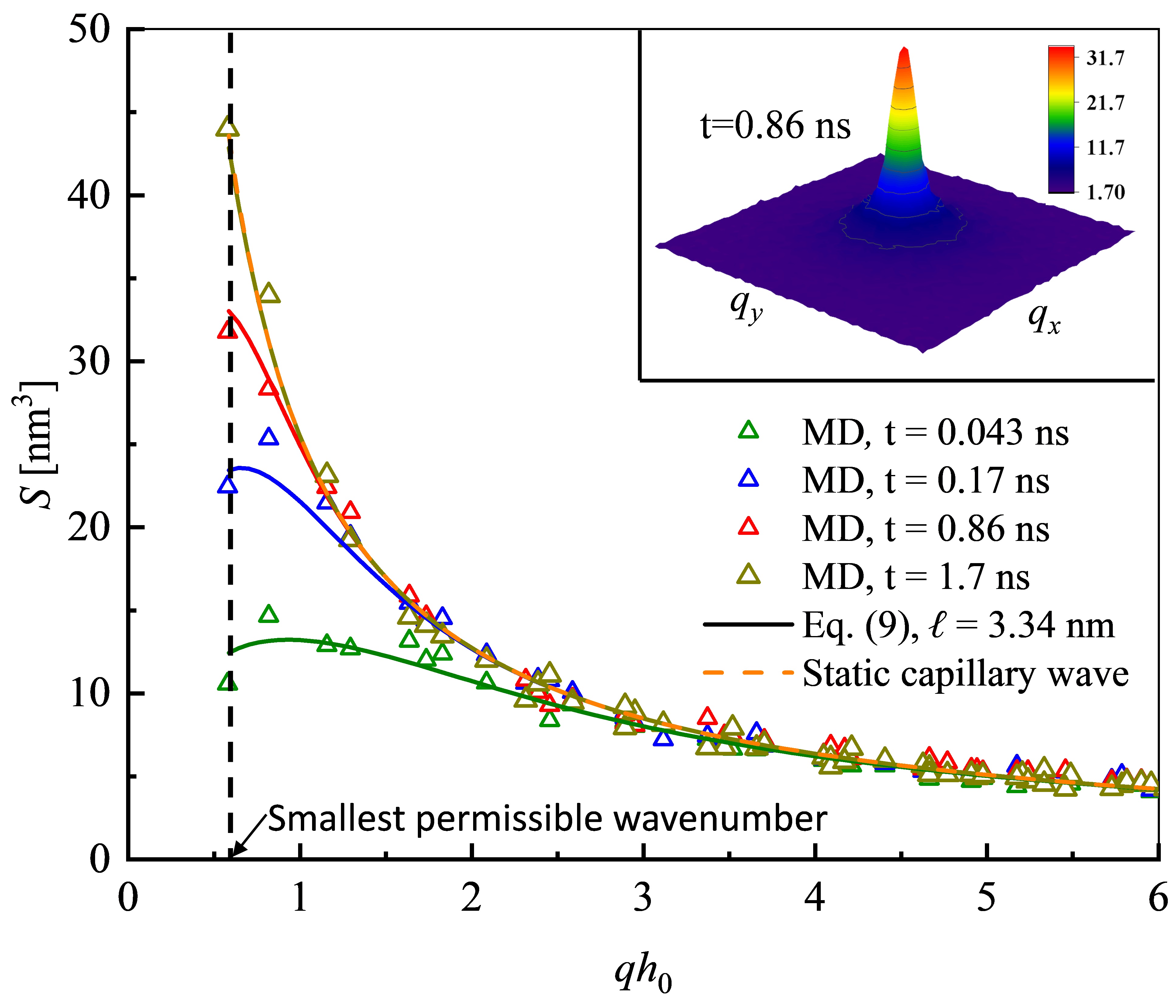}
\caption{Capillary wave growth for the isotropic case $C_{\left\langle 100\right\rangle,\,2}$. MD spectra after averaging over azimuthal directions are shown by triangles, with different colors represent different time instances. The analytical solutions are given by solid lines. The inset shows the 2D spectra from the 2D Fourier transform of the free surface at $t=0.86$ ns.}  \label{fig3}
\end{figure}
The growth of capillary waves to the static spectrum, from an initially smooth interface, is shown in Figure \ref{fig3} for case $C_{\left\langle 100\right\rangle,\,2}$ where slip length $\ell_x=\ell_y=3.34$ nm. As the simulation starts, the free surface becomes rough, and the Fourier transform of the free surface shape is performed at various intervals to obtain the evolving spectra. As shown in the inset of Figure \ref{fig3}, the spectra of surface waves from MD simulations is radially symmetric, which is expected from eq.\,5 and eq.\,\ref{dis} due to the isotropic slip length. Thus, the spectra can be averaged over the azimuthal direction and represented in terms of $q=\sqrt{q_x^2+q_y^2}$, with the results shown by triangles in Figure \ref{fig3}. The spectra are also ensemble averaged over $40$ independent realizations. It can be seen that the spectra evolve with time but the static spectrum given by the capillary wave theory forms an upper limit. The transition time for the smooth surface to reach the static spectrum is predicted to be $1/\Omega(q=q_{min})$, where $q_{min}=2\pi/L$ is based on the longest wavelength we can find on the periodic surface, and it is evaluated to be about $t_s=1$ ns. This is confirmed by the positions of the dark yellow symbols in Figure \ref{fig3}, at time $t=1.7$ ns, which show the surface has safely reached the static spectrum, whereas at $t=0.86$ ns (red symbols) the amplitude of the smallest wavenumber does not reach the value predicted by the static spectrum.

 To measure the correlation of capillary waves presented in the next subsections, it is important to make sure the surface has the state of thermal equilibrium characterised by the static spectrum. As shown, the transition time found from our knowledge of the growth of capillary waves provides a useful guideline. In the long-wave approximation, $2\pi h_0/L \ll 1$, one can find the transition time scales with $L^4$ from the Stokes dispersion relation\cite{zhang2020Thermal}, which means that care should be taken when interpreting results for larger film lengths where reaching thermal equilibrium (the static spectrum) for the surface is often computationally intractable for molecular simulations. This problem seems to appear in previous MD simulations of thermal capillary waves using very long films\cite{wi2010}. It may also be relevant in experimental studies of thermal capillary waves using high-viscosity polymers, where the film length is at the microscale or macroscale\cite{zhang2020Thermal}. 
\subsection{Relaxation of capillary wave correlations with varying slip}
\begin{figure}[t!]
\includegraphics[scale=0.1]{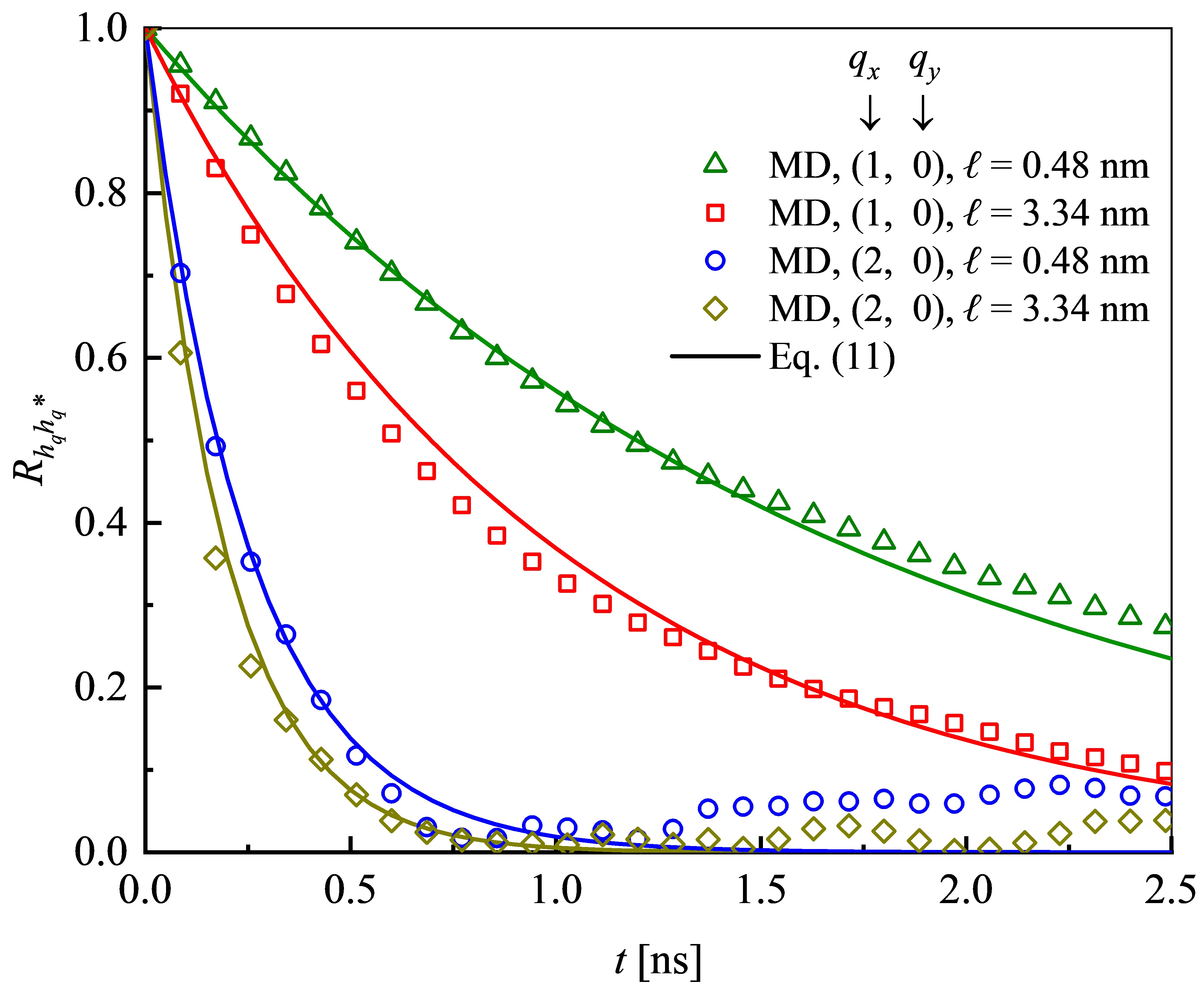}
\caption{Effects of varying slip length on temporal correlations of thermal capillary waves. MD results (different symbols) are compared with the analytical solutions eq. \ref{eq6} (solid lines). Wave vectors are represented by $\bm{q}=(q_x,q_y)=\frac{2\pi}{L}(n_x,n_y)$, where $(n_x,n_y)$ are pairwise integers, and without loss of generality, we consider waves in the $x$-direction due to the isotropy.}\label{fig4}
\end{figure}
In this subsection, we explore the effect of different slip lengths on the relaxation of capillary wave correlations using the isotropic $\left\langle 100 \right\rangle$ surface. As shown earlier, increased slip length is achieved by decreasing the liquid-solid interactions; case $C_{\left\langle 100\right\rangle,\,2}$ has slip length $\ell=3.34$ nm, while case $C_{\left\langle 100\right\rangle,\,1}$ has slip length $\ell=0.48$ nm. After the transition time of each case ($t_s=1.7$ ns for case $C_{\left\langle 100\right\rangle,\,1}$ and $t_s=1.0$ ns for case $C_{\left\langle 100\right\rangle,\,2}$), the 2D Fourier transform of the surface position at different time instances is performed. Temporal correlations of interfacial Fourier modes, which depend on the time interval $|t-t'|$, are then calculated in a standard way and averaged over 10,000 times for each time interval. Wave vectors are represented by $\bm{q}=(q_x,q_y)=\frac{2\pi}{L}(n_x,n_y)$ where $(n_x,n_y)$ are pairwise integers, characterising the wave numbers. As the relaxation of capillary wave correlations on the $\left\langle 100\right\rangle$ surface is expected to be radially symmetric (see further discussions in the next subsection), only correlations at $x$-direction ($q_y=0$) are presented in Figure \ref{fig4} without loss of generality. In Figure \ref{fig4}, for the same wave vector, it can be seen that larger slip length leads to faster decay of $R_{h_qh_q^*}$, from the MD results (green triangles and red squares, for instance). On the other hand, given the same slip length, wave vectors with larger wavenumbers (the norm of the wave vector) decay faster (see green triangles and blue circles, for instance). Both of these features, and the actual values, are well predicted by the Langevin equation results from  eq \ref{eq6} and eq \ref{dis}, using the independently measured slip length. 
\begin{figure}[th!]
\includegraphics[width=\linewidth]{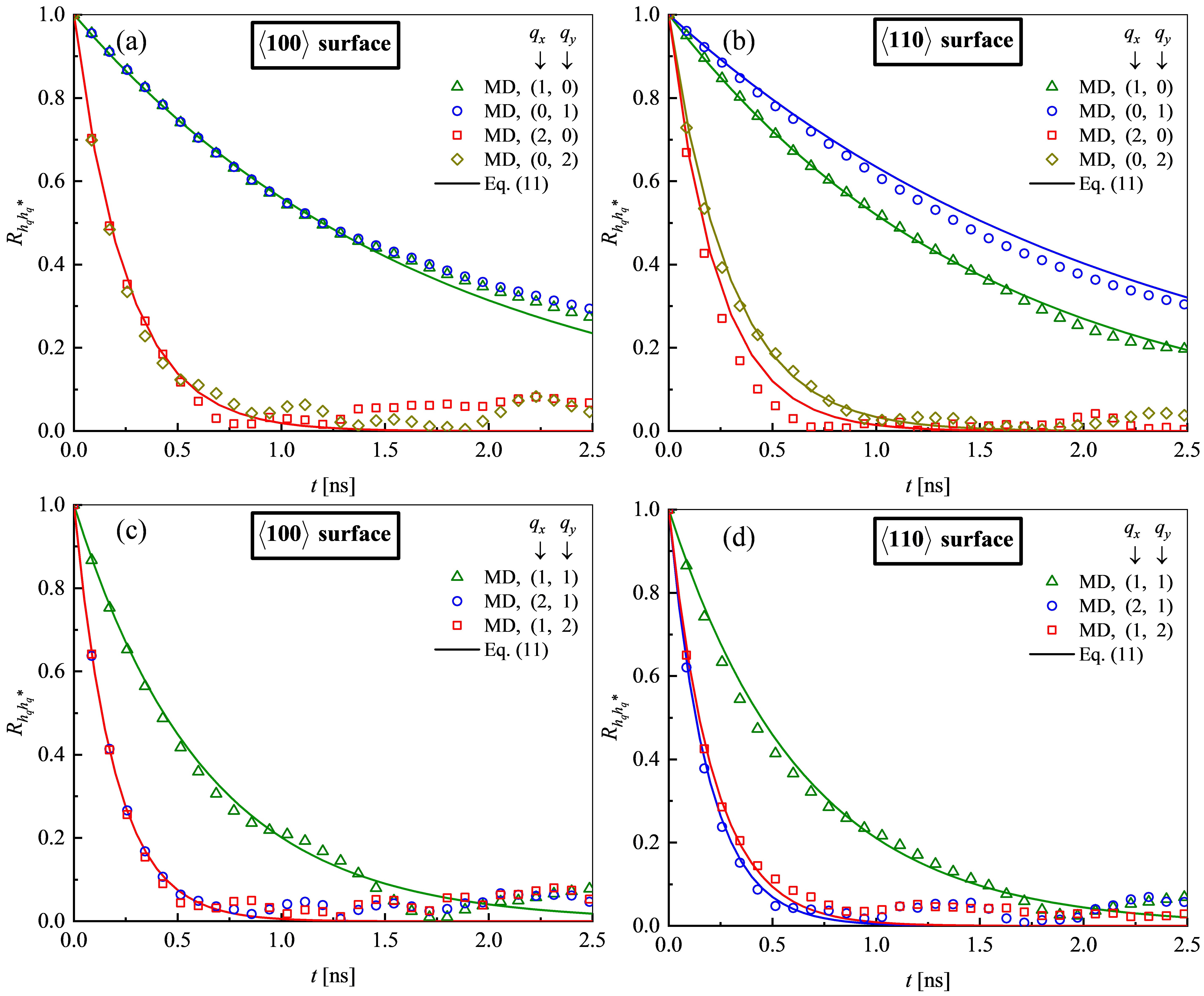}
\caption{Effects of anisotropic slip on temporal correlations of thermal capillary waves. MD results (symbols) are compared with analytical solutions (solid lines). Figures (a) and (c) are for the isotropic $\left\langle 100\right\rangle$ surface, and figures (b) and (d) are for the anisotropic $\left\langle 110\right\rangle$ surface. Wave vectors are represented by $(q_x,q_y)=\frac{2\pi}{L}(n_x,n_y)$ where $(n_x,n_y)$ are pairwise integers.}  \label{fig5}
\end{figure}
\subsection{Relaxation of capillary wave correlations with anisotropic slip }
Consider now the effects of anisotropic slip on the relaxation of capillary wave correlations. Figure \ref{fig5} shows the comparison of wave correlations with different orientations for the  case $C_{\left\langle 100\right\rangle,\,1}$ and case $C_{\left\langle 110\right\rangle}$. For the isotropic $\left\langle 100 \right\rangle$ surface, the relaxation of correlations of wave vectors at different directions for a given wavenumber are the same as shown in Figures \ref{fig5}(a) and (c). For example, the relaxation of the correlation of a wave vector at the direction (1, 0) is the same as that of a wave vector at (0, 1), and the relaxation of the correlation of a wave vector at (1, 2) is the same as that of a wave vector at (2, 1). However, for the anisotropic $\left\langle 110 \right\rangle$ surface, as shown in Figures \ref{fig5} (b) and (d), the relaxation of correlations of waves for a given wavenumber varies with direction. For example, the correlation of the $(1, 0)$ wave vector relaxes faster than that of the wave vector $(0, 1)$. The wave vector $(2, 1)$ also  decays faster than the wave vector $(1, 2)$ in terms of correlations. Clearly, the anisotropic relaxation of capillary wave correlations is due to the anisotropic-slip boundary condition. A wave vector closer to the $x$-direction having a faster decay than a wave vector (with the same wavenumber) closer to the $y$-direction; this is also due to the larger slip length in the $x$-direction similar to the case in the previous subsection. However, the difference becomes smaller when the wavenumber is increased. This means that to infer the anisotropy of a substrate from measuring the correlations of capillary waves, it is better to measure the correlations of waves with smaller wavenumbers. Using the measured slip length, and the derived new dispersion relation, the relaxation of capillary waves for $\left\langle 110\right\rangle$ surface can be predicted well by eq \ref{eq6} and eq \ref{dis} (see solid lines). 

\begin{figure}[th!]
\includegraphics[scale=0.1]{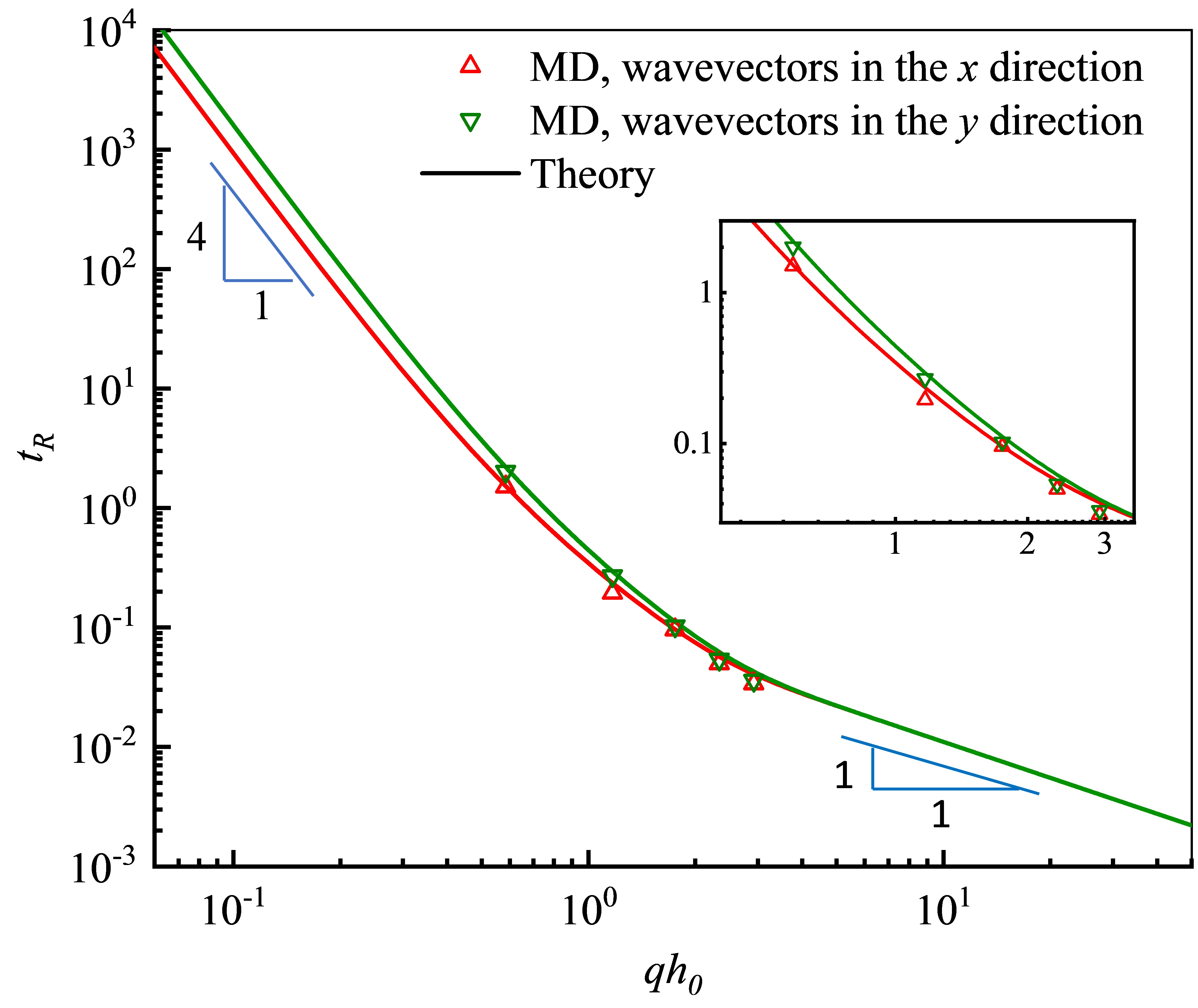}
\caption{Relaxation time as a function of wavenumbers. MD results (symbols) are compared with analytical solutions (solid lines).}  \label{fig6}
\end{figure}

Now we consider the relaxation time $t_R$, which is equal to $1/\Omega$. We focus on the relaxation time of wave vectors along the $x$ or $y$ direction, as the relaxation of those waves only depends on the slip length in that direction. Figure \ref{fig6} shows the values of relaxation time obtained from MD simulations agree well with the analytical solutions. A simple asymptotic analysis of the dispersion in different limits of $qh_0$, shows that there are two scaling relations between the dispersion relation and non-dimensional wavenumber $qh_0$.  
 Firstly, for a very long film (or thin film) $qh_0\ll 1$ with small slip $\ell_x \le h_0$ and $\ell_y \le h_0$, the dispersion relation eq \ref{dis} can be significantly simplified to (see Supporting Information)
\begin{equation}
\Omega(q_x,q_y)\approx \frac{\gamma}{3\mu}q^4h_0^2\left[h_0+3\frac{\left(q_x^2\ell_x+q_y^2\ell_y\right)}{q^2}\right].
\end{equation}
Thus, the relaxation time $t_R$ scales with $q^{-4}$ for a wave vector along the $x$ or $y$ direction, which is supported by Figure \ref{fig6} when $qh_0$ becomes small. This also means that the relaxation time for $q=2\pi/L$ scales with $L^4$ so that it becomes experimentally measurable for long films (the relaxation time for a short film is a few nanoseconds in our MD simulations).

 Secondly, if $qh_0\gg 1$, the dispersion relation eq \ref{dis} is simplified to 
\begin{equation}\label{eq25}
\Omega(q_x,q_y)\approx\frac{\gamma}{2\mu}q,
\end{equation}
so that the relaxation time scales with $q^{-1}$, which is shown by Figure \ref{fig6} when $qh_0$ goes to infinity. Equation \ref{eq25} also shows that the dispersion relation becomes independent of slip length, as we may expect for thick films ($qh_0\gg 1$), and becomes isotropic so that we can not use thermal capillary waves in this regime to measure slip length. In particular, the temporal correlations shown in Figure 4 for different slip length (but with the same wavenumber) will overlap each other so that the effects of slip on the relaxation of correlations will not be observed.

We note that if the slip length is very large $\ell \gg h_0$, there may be other scaling relations, as shown in ref \cite{he2007}, though the analysis there is for isotropic slip. 
\section{Conclusion}\label{sec6}
In this work, the effects of anisotropic slip on the relaxation of correlations of thermal capillary waves are investigated both numerically and analytically. We perform molecular dynamics simulations of liquid films bounded by isotropic-slip substrates and anisotropic-slip substrates. The correlations of Fourier modes obtained from simulations compare well with a Langevin equation, where a new dispersion relation considering the anisotropic-slip boundary condition is derived. Our results show that the larger slip length leads to faster decay of the correlations, and the anisotropic-slip leads to anisotropic relaxations of capillary wave correlations. Notably, other factors besides the anisotropic slip found here may lead to an anisotropic relaxation of capillary waves as well, such as the orientation dependent stiffness coefficient\cite{hoyt2002atomistic,benet2014computer}.

Though the anisotropic surface used in our MD simulations is ideal, it may provide inspiration for making engineered surfaces that are anisotropic in slip, which may be useful in micro- or nano-fluidics to obtain a directional control on liquid transport. We also believe this work strengthens the applicability of using thermal capillary waves as a non-invasive method to infer slip length in future experiments using existing isotropic or anisotropic-slip substrates. The temporal correlations can be directly obtained in experiments using x-ray techniques\cite{po2015,ki2003} and they can be fitted with analytical solutions developed here to obtain slip length, which is the usual way to infer properties of interests using surface waves. Notably, this article provides guidance on the experimental setup, in particular showing that slip length is not measurable from temporal correlations of capillary waves of thick films ($qh_0\gg 1$).

\begin{acknowledgement}
We wish to acknowledge the financial support by the EPSRC (grants EP/N016602/1, EP/P020887/1, EP/S029966/1 \& EP/P031684/1).
\end{acknowledgement}

\begin{suppinfo}
The following file is available free of charge.
\begin{itemize}
  \item SupportingInformation.pdf: a derivation of the anisotropic dispersion relation.
\end{itemize}

\end{suppinfo}

\providecommand{\latin}[1]{#1}
\makeatletter
\providecommand{\doi}
  {\begingroup\let\do\@makeother\dospecials
  \catcode`\{=1 \catcode`\}=2 \doi@aux}
\providecommand{\doi@aux}[1]{\endgroup\texttt{#1}}
\makeatother
\providecommand*\mcitethebibliography{\thebibliography}
\csname @ifundefined\endcsname{endmcitethebibliography}
  {\let\endmcitethebibliography\endthebibliography}{}

\end{document}